\documentclass[preprint,showpacs,preprintnumbers,amsmath,amssymb,superscriptaddress]{revtex4}
\usepackage{graphicx}
\usepackage{bm}
\usepackage{color,graphics}

\begin{document}

\title{Connection between the two branches of the quantum two-stream instability across the $\textbf{k}$  space}

\author{A. Bret}
\affiliation{ETSI Industriales, Universidad de Castilla-La Mancha, 13071 Ciudad Real, Spain}

\author{F. Haas}
\affiliation{Universidade do Vale do Rio dos Sinos, Av. Unisinos 950, 93022-000 S\~ao Leopoldo RS, Brazil}

\begin{abstract}
The stability of two quantum counter-streaming electron beams is investigated within the quantum plasma fluid equations for arbitrarily oriented wave vectors $\mathbf{k}$. The analysis reveals that the two quantum two-stream unstable branches are indeed connected by a continuum of unstable modes with oblique wave vectors. Using the longitudinal approximation, the stability domain for any $\mathbf{k}$ is analytically explained, together with the growth rate.
\end{abstract}


\maketitle

\section{Introduction}
The theory of quantum plasma is currently rapidly expanding as a result of its potential applications ranging from micro-electronic \cite{Markowich} to laser-plasma interaction \cite{Kremp1999,Kull2001} or astrophysics \cite{Chabrier2002,ApJQuantum}. Since the derivation of the quantum fluid equations \cite{Haas2000,Manfredi2001}, many well known classical plasmas effects including plasmas oscillations with or without magnetic field \cite{HaasGarcia2003,Khann2009}, plasma instabilities \cite{Haas2000,BretPoPQuantum2007,BretPoPQuantum2008,HaasWeibel2008}, solitons and vortices \cite{ShuklaSolitons}, or turbulence \cite{haasturbulence}, have been re-examined in their quantum version.

Beyond their potential applications, many of these problems present a fundamental interest as they form the basis for the general theory of quantum plasmas. Among them, the quantum version of the two-stream instability is endorsing the r\^ole its classical counterpart plays with classical plasma: that of a fundamental process worthy of investigation in itself, and included in every plasma physics textbook. Since the first works on the quantum two-stream instability, it has been recognized that one of the most striking difference between the quantum and the classical treatment of this process is that the quantum version displays two unstable branches instead of only one for the classical case \cite{Haas2000}. Kinetic treatment are now available showing that thermal spread can suppress both quantum branches \cite{Anderson2002}, as they do with the single unstable branch in the classical case.

While unstable quantum modes in counter-streaming beams systems have been assessed for modes with wave vector aligned (two-stream) or perpendicular (filamentation \cite{BretPoPQuantum2007,BretPoPQuantum2008}) with the streams, intermediate orientations have not been considered yet. The intent of this paper is precisely to fill this gap. As a result, an interesting conclusion is reached: the two branches of the quantum two-stream instability are indeed connected with each other. By letting the wave vector acquire a perpendicular component, we found a continuum of unstable modes bridging the two quantum branches across the $\textbf{k}$ space.

\section{Dispersion equation for arbitrary modes}
We here derive the dispersion equation for arbitrarily oriented wave vector. For the sake of simplicity, we deal with the generic system formed by two cold counter-streaming electron beams of equal density $n_0/2$ and opposite velocities $\pm v_0\hat{\mathbf{z}}$. Ions form a fixed neutralizing background with density $n_i=n_0$ and there's no background magnetic field. The dispersion equation is derived from Maxwell's equations and from the quantum fluid equations for the two electrons species,
\begin{equation}\label{eq:conser}
   \frac{\partial n_\pm }{\partial t}+\nabla\cdot(n_\pm  \mathbf{v}_\pm )=0,
\end{equation}
and,
\begin{equation}\label{eq:force}
   \frac{\partial \mathbf{v}_\pm }{\partial
   t}+(\mathbf{v}_\pm \cdot\nabla)\mathbf{v}_\pm =-\frac{q}{m}\left(\mathbf{E} + \frac{\mathbf{v}_\pm \times
   \mathbf{B}}{c}\right)+\frac{\hbar^2}{2
   m^2}\nabla\left(\frac{\nabla^2\sqrt{n_\pm }}{\sqrt{n_\pm }}\right),
\end{equation}
where $-q<0$ and $m$ are the electron charge and mass respectively. The subscripts ``+'' and ``-'' stand here for the electron population going to the positive and negative directions respectively. These equations are now linearized perturbing every quantity with a term proportional to $\exp(i \mathbf{k}\cdot\mathbf{r} - i\omega t)$, where $\mathbf{k}=(k_x,0,k_z)$. The linearized mass conservation equation yields straightforwardly,
\begin{equation}\label{eq:conserline}
    n_{\pm 1}=n_{\pm 0}\frac{\mathbf{k}\cdot \mathbf{v}_{\pm 1}}{\omega-\mathbf{k}\cdot
    \mathbf{v}_{\pm 0}},
\end{equation}
while the linearized quantum Euler equation reads,
\begin{equation}\label{eq:forceL}
 i  (\mathbf{k}\cdot \mathbf{v}_{\pm 0}-\omega) \mathbf{v}_{\pm 1}
  = -\frac{q}{m}\left(\mathbf{E}_{1}+\frac{\mathbf{v}_{\pm 0}\times
   \mathbf{B}_1}{c}\right)-i\frac{\hbar^2 k^2}{4m^2}\frac{n_{\pm 1}}{n_{\pm 0}}\mathbf{k}.
\end{equation}
While flow aligned unstable modes ($k_x=0$) are exactly electrostatic with $\mathbf{B}_1=0$,  such is not the case for a general orientation of the wave vector. This is why the exact calculation requires a Lorentz force term. Although some approximations will be used in the sequel to analyze the problem, we first outline the exact one.

\begin{figure}[t]
\begin{center}
\includegraphics[width=0.4\textwidth]{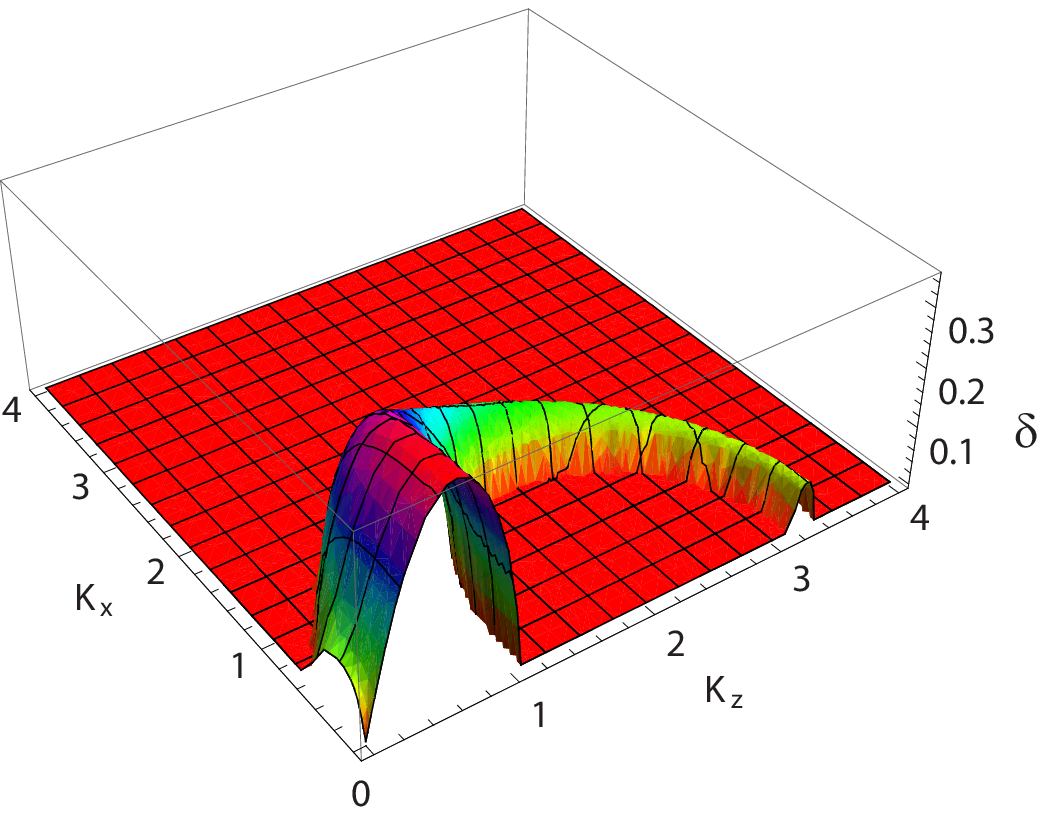}\\\includegraphics[width=0.4\textwidth]{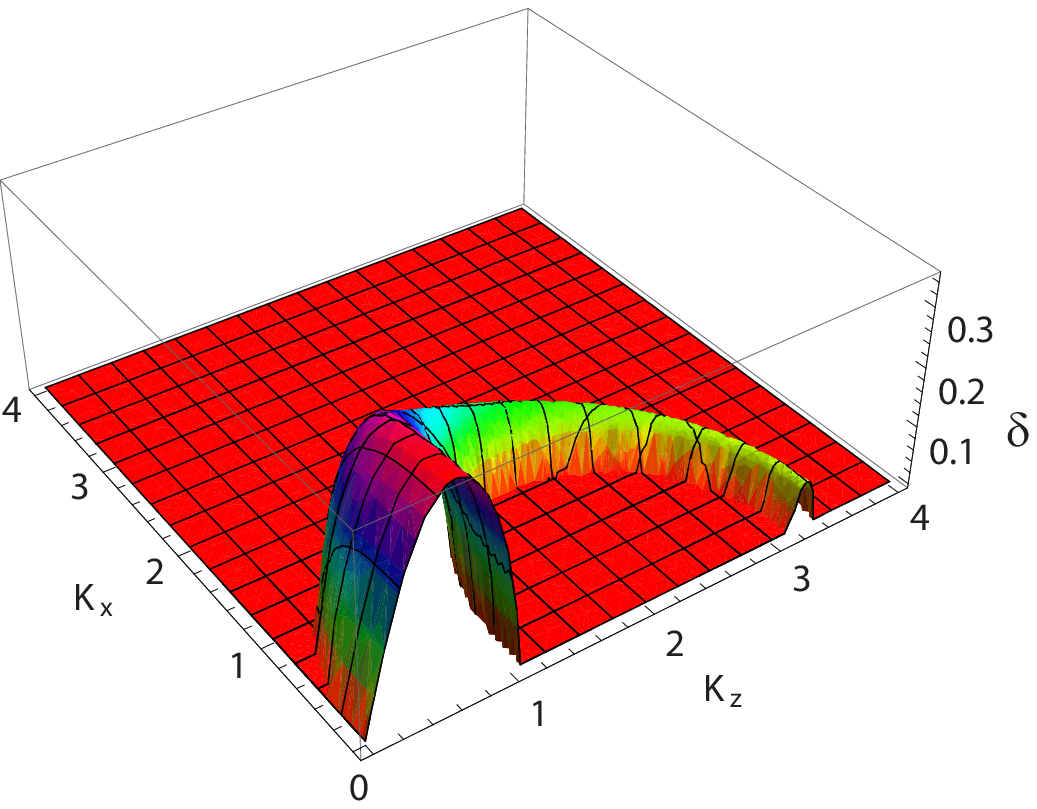}
\end{center}
\caption{Growth rate $\delta$ in terms of $\mathbf{K}$ for $v_0/c=0.1$ and $H=0.6$. Top: Exact calculation. Bottom: Longitudinal approximation.} \label{fig:1}
\end{figure}

The linearized equation (\ref{eq:conserline}--\ref{eq:forceL}) allow to express the perturbed velocities $\mathbf{v}_{\pm 1}$ in terms of the perturbed fields $\mathbf{E}_{1}$ and $\mathbf{B}_{1}$. The first order magnetic field is then eliminated through the Maxwell-Faraday equation $\mathbf{B}_{1}=(c/\omega)\mathbf{k}\times\mathbf{E}_{1}$ so that the first order current $\mathbf{J}_1= -q \sum_\pm (n_{\pm 0}\mathbf{v}_{\pm 1}+n_{\pm 1}\mathbf{v}_{\pm 0})$, can eventually be expressed in terms of $\mathbf{E}_{1}$ only. Combining now Maxwell-Faraday and Maxwell-Amp\`{e}re equations gives,
\begin{equation}\label{eq:combimaxwell}
\frac{c^2}{\omega^2}\mathbf{k}\times(\mathbf{k}\times\mathbf{E}_1)+\mathbf{E}_1+\frac{4i\pi}{\omega}\mathbf{J}_1=0,
\end{equation}
from which the formal dispersion relation is finally derived,
\begin{equation}\label{eq:disperformal}
\frac{c^2}{\omega^2}\mathbf{k}\times(\mathbf{k}\times\mathbf{E}_1)+\epsilon(\mathbf{k},\omega)\cdot\mathbf{E}_{1}=0,
\end{equation}
where $\mathbf{\epsilon}(\mathbf{k},\omega)=\mathbf{1}+\frac{4i\pi}{\omega}\mathbf{J}_1$ is the dielectric tensor. The dispersion equation derived from Eq. (\ref{eq:disperformal}) is very involved and will not be reproduced here. Instead, we now introduce the very simple longitudinal approximation which, in the classical case, has been found worthy over the full wave-vectors spectrum except near the perpendicular direction \cite{BretPRE2004,BretPRE2005}. The dispersion equation for longitudinal waves reads \cite{Melrose2009},
\begin{equation}\label{eq:disperlongi}
    0=1-\sum_{\pm}\frac{4\pi q^2n_0/2}{ m}\int \frac{f_{\pm}(\mathbf{v}) d^3\mathbf{v}}{(\omega-\mathbf{k}\cdot \mathbf{v})^2-(\hbar k^2/2m)^2},
\end{equation}
where $f_\pm$ is the distribution function of each beam. With $f_\pm(\mathbf{v})=\delta(v_x)\delta(v_y)\delta(v_z\mp v_0)$, the result reads $\epsilon(\Omega)=1-F(\Omega)=0$, where
\begin{equation}\label{eq:longi}
    F(\Omega)=\frac{1}{2} \left[\frac{1}{(\Omega -K_z)^2-\frac{1}{4} H^2 \left(K_x^2+K_z^2\right)^2}+\frac{1}{(\Omega +K_z)^2-\frac{1}{4} H^2 \left(K_x^2+K_z^2\right)^2}\right],
\end{equation}
in terms of the non-dimensional parameters,
\begin{equation}\label{eq:param}
    \mathbf{K}=\frac{\mathbf{k}v_0}{\omega_p},~~ \Omega=\frac{\omega}{\omega_p},~~  H=\frac{\hbar\omega_p}{mv_0^2},
\end{equation}
where $\omega_p=4\pi n_0q^2/m$ is the electron plasma frequency (note that the plasma frequency is here defined for a density $n_0$ different from the beam electronic densities $n_0/2$). The rescaled growth rate of the instability $\delta$ in terms of the rescaled wave vector $\mathbf{K}$ is plotted on Figure \ref{fig:1} without any approximation (top), and according to the longitudinal approximation (bottom). As expected, the exact calculation displays unstable filamentation modes for $K_z=0$ which are bypassed by the approximation. Apart for the quasi normal wave vectors, the approximation is very good, and even exact for $K_x=0$. The most striking features here is the connection of the two two-stream branches along the flow axis $K_z$. When restricting the calculation to this axis, these two branches seem disconnected. After extending the analysis for arbitrarily oriented wave vectors, we now see that there is a continuum of unstable modes bridging between them.

\begin{figure}[t]
\begin{center}
\includegraphics[width=0.4\textwidth]{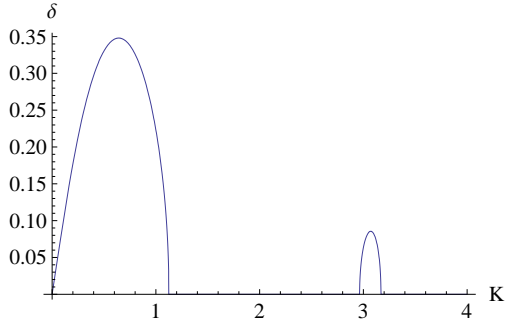}
\end{center}
\caption{Growth rate $\delta$ as a function of $K$ for $H=0.6$ and $\theta=\pi/6$.} \label{fig:2}
\end{figure}

\section{Connection between the two-stream modes and unstable domain}
\begin{figure*}[t]
\begin{center}
\includegraphics[width=0.3\textwidth]{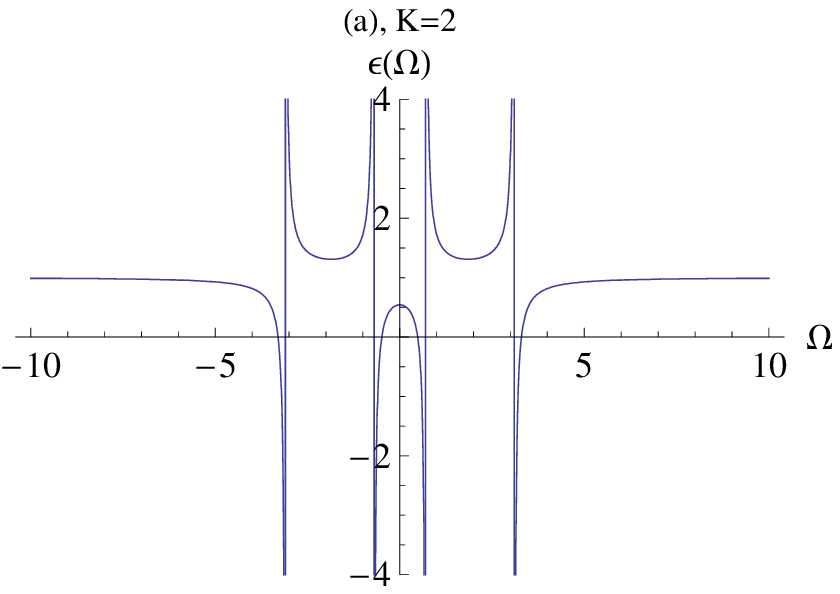}
\includegraphics[width=0.3\textwidth]{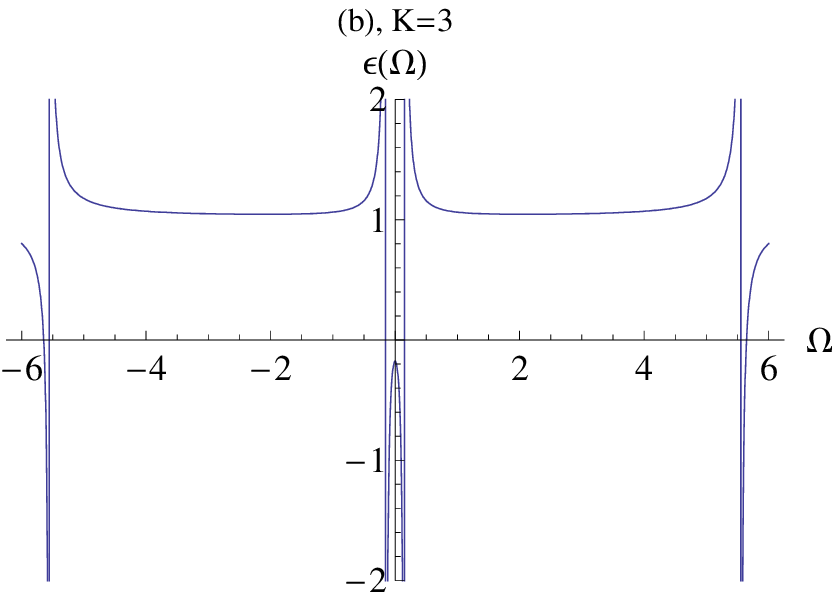}
\includegraphics[width=0.3\textwidth]{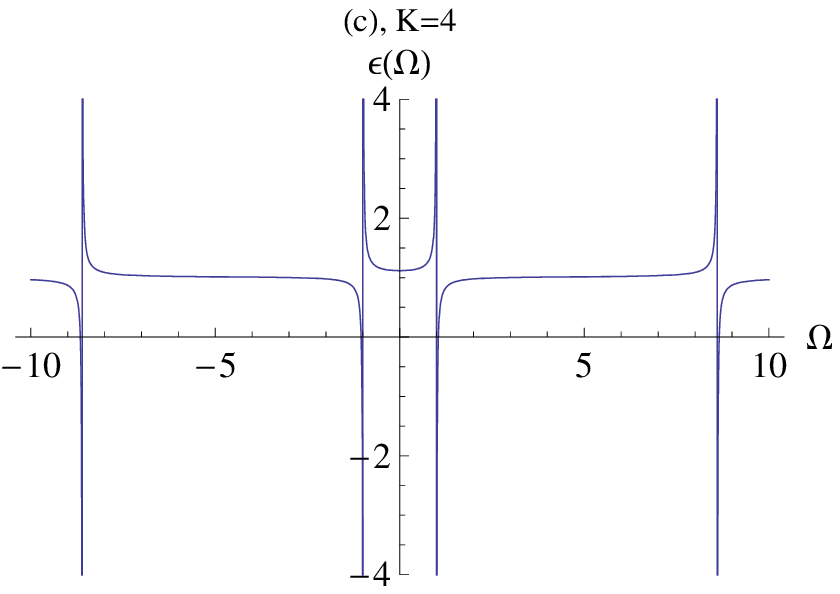}
\end{center}
\caption{Dispersion function according to Eq. (\ref{eq:longi})  $H=0.6$, $\theta=\pi/6$ and $K=2$ (a), $K=3$ (b) and $K$=4 (c). } \label{fig:3}
\end{figure*}

An analytical analysis of the dispersion function for $(K_x,K_z)\neq (0,0)$ is possible and allows for an analytical characterization of the bridge between the two-stream branches. For the sake of the discussion, we switch to polar coordinates,
\begin{eqnarray}\label{eq:polar}
K_x\equiv K\sin\theta,\nonumber\\
K_z\equiv  K\cos\theta,
\end{eqnarray}
and consider $\theta=\pi/6$. Figure \ref{fig:2} pictures the growth rate as a function of $K$ in this direction. Figures \ref{fig:3} now display the behavior of the dispersion function for $K=2$ (stable), $K=3$ (unstable) and $K=4$ (stable). The stability of the system can be explained from the respective position of the singularities of the dispersion function. According to Eq. (\ref{eq:longi}), these singularities read,
\begin{eqnarray}\label{eq:singu}
\Omega_1&=&- Z \cos \theta -\frac{1}{2}H K^2,\nonumber\\
\Omega_2&=&- Z \cos \theta +\frac{1}{2}H K^2,\nonumber\\
\Omega_3&=& Z \cos \theta -\frac{1}{2}H K^2,\nonumber\\
\Omega_4&=& Z \cos \theta +\frac{1}{2}H K^2.
\end{eqnarray}

For $K=2$ in Fig. \ref{fig:3}a, we have $\Omega_1<\Omega_2<\Omega_3<\Omega_4$. The system is here stable because  $\epsilon(\Omega)>0$, which gives 4 real roots to the 4th degree dispersion equation. For $K=3$, the ordering of the singularities is still the same, but $\epsilon(0)<0$, rendering the system unstable. Finally, for $K=4$, the ordering is $\Omega_1<\Omega_3<\Omega_2<\Omega_4$, namely, $\Omega_3$ and $\Omega_2$ switched their positions. The stability in this case is necessary from the behavior of the dispersion function near its singularities. When their order is the one of Fig. \ref{fig:3}c, i.e. ``1,3,2,4'', the dispersion equation necessarily crosses the $\Omega$ axis 4 times and the 4th degree dispersion equation has therefore only purely real roots. If the ordering is ``1,2,3,4'', like the one found for $K=2$ or 3, stability relies on the sign of $F(0)$. From Figs. \ref{fig:3}a to \ref{fig:3}b, the system is therefore unstable until $\Omega_2=\Omega_3$ which gives,
\begin{equation}\label{eq:Kstab}
    K^*=\frac{2}{H}\cos\theta,
\end{equation}
defining thus the largest unstable $K$ in a given direction. For $\theta=0$, the quantity $2/H$ is exactly the upper limit of the large $K$ instability two-stream branch \cite{Haas2000}. For an arbitrary angle, the equation above describes a circle of diameter $2/H$ and center $(K_x,K_z)=(0,1/H)$ which fits perfectly as is observed on Figs. \ref{fig:1}. Finally, the expression of $K^*$ breaks down for $\theta\sim\pi/2$; according to Eq. (\ref{eq:Kstab}), $K^*(\pi/2)=0$ while the fluid quantum theory of the filamentation instability gives a finite value for this quantity \cite{BretPoPQuantum2007}.

\begin{figure}[t]
\begin{center}
\includegraphics[width=0.4\textwidth]{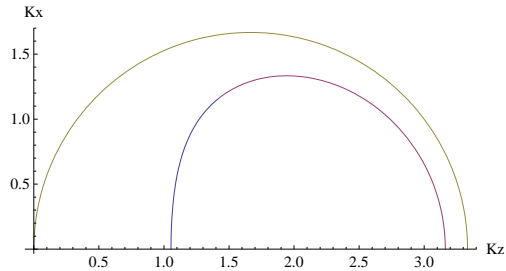}
\end{center}
\caption{Plot of $K^*$ (yellow), $K_-$ (blue) and $K_+$ (purple) given by Eqs. (\ref{eq:Kstab}--\ref{eq:Kstab2}). $K_-=K_+$ for $\theta=\arccos\sqrt{H}$. Wave vectors between the two frontiers are unstable.} \label{fig:4}
\end{figure}

Turning now the sign of $\epsilon(0)$, a little algebra gives $\epsilon(0)=0$ for,
\begin{eqnarray}
    K_-(\theta)&=&\frac{\sqrt{2}}{H}\sqrt{\cos\theta^2-\sqrt{\cos\theta^4-H^2}},~~\theta\in [0,\arccos\sqrt{H}],\nonumber\\ \label{eq:Kstab2}
    \mathrm{or}, \\ 
    K_+(\theta)&=&\frac{\sqrt{2}}{H}\sqrt{\cos\theta^2+\sqrt{\cos\theta^4-H^2}},~~\theta\in [0,\arccos\sqrt{H}]. \nonumber
\end{eqnarray}
From $\theta=0$ to $\arccos\sqrt{H}$, $K_+(\theta)$ is monotonically  decreasing, and systems with $K<K_+(\theta)$ are stable. In the same interval, $K_-(\theta)$ is monotonically  increasing, and systems with $K<K_-(\theta)$ are unstable. The unstable domain defined by Eqs. (\ref{eq:Kstab}--\ref{eq:Kstab2}) is graphically summarized in Figure \ref{fig:4} when one recognizes immediately the domain outlined in Figs. \ref{fig:1}.

\section{Growth rate analysis}
It turns out that the dispersion equation from Eq. (\ref{eq:longi}) is biquadratic and can be solved exactly for any parameter set. In terms of $\Omega^2$, the two roots are,
\begin{eqnarray}\label{eq:growthrateanal}
\Omega_1^2&=&\frac{1}{4} \left[2+4 K_z^2+H^2 \left(K_x^2+K_z^2\right)^2-2\sqrt{1+4 K_z^2 \left[2+H^2 \left(K_x^2+K_z^2\right)^2\right]}\right],\nonumber\\
\Omega_2^2&=&\frac{1}{4} \left[2+4 K_z^2+H^2 \left(K_x^2+K_z^2\right)^2+2\sqrt{1+4 K_z^2 \left[2+H^2 \left(K_x^2+K_z^2\right)^2\right]}\right].
\end{eqnarray}
Unstable modes clearly arises from $\Omega_1^2$, since $\Omega_2^2>0$ for any parameters. Of course, the sign analysis of $\Omega_1^2$ yields the unstable domain already explained.

Analytic results are acessible not only in the electrostatic approximation. Indeed, allowing for magnetic field perturbations and following the steps outlined in Section II, the exact growth rate $\delta$ can be found for the filamentation instability ($K_z = 0$), without any further restriction, as 
\begin{equation}
\label{fil}
\delta^2 = \frac{1}{2}\left(\left[\left(1 + \frac{c^2 K_{x}^2}{v_{0}^2} + \frac{H^2 K_{x}^4}{4}\right)^2 +  K_{x}^2 \left(4 - H^2 K_{x}^2 (1 + \frac{c^2 K_{x}^2}{v_{0}^2})\right)\right]^{1/2} - 1 - \frac{c^2 K_{x}^2}{v_{0}^2} - \frac{H^2 K_{x}^4}{4}\right) \,.
\end{equation}
From Eq. (\ref{fil}) it can be shown that sufficiently large wave-numbers such that 
\begin{equation}
\frac{H^2 K_{x}^2}{4}\left(1 + \frac{c^2 K_{x}^2}{v_{0}^2}\right) \geq 1
\end{equation}
are stable with respect to the filamentation instability. This stabilization effect can exist only in the quantum case ($H \neq 0$).

\section{Conclusion}
The essential results of this work are in Fig. 1, where the apparently distinct unstable quantum two-stream modes are shown to be connected. This achievement follows after considering the two-dimensional ${\bf k}$-space and including the Lorentz-force term in the pertinent quantum hydrodynamic equations. For arbitrary wave-vector orientation, exact analytic calculations are possible in the electrostatic approximation, which is fairly reasonable except for wave propagation near the axis normal to the beams direction. In this later (filamentation) case, the exact growth rate is also acessible. In the general electromagnetic case, the quantum filamentation and quantum two-stream instabilities are viewed as a single, unified phenomenon.

\begin{acknowledgments}
This work has been  achieved under projects FIS 2006-05389 of the
Spanish Ministerio de Educaci\'{o}n y Ciencia and PAI08-0182-3162 of
the Consejer\'{i}a de Educaci\'{o}n y Ciencia de la Junta de
Comunidades de Castilla-La Mancha.
\end{acknowledgments}

\end{document}